\documentstyle[preprint,aps]{revtex}
\begin{document}
 \font\script=cmr10 at 12 pt
\newcommand{\qtwo}{\stackrel{II}{q}}
 \newcommand{\pitwo}{\stackrel{II}{\Pi}}

\preprint{NCL95-TP1}
\tighten
\title{STOCHASTIC DYNAMICS OF LARGE-SCALE INFLATION \\
 IN DE~SITTER SPACE}

\author{O.~E.~Buryak}
 \address{Keldysh Institute of Applied Mathematics \\
Russian Academy of Sciences, Miusskaya Sq. 4 \\
Moscow 125047, Russia \\
 {\rm and}\\
 Dept.~of
Physics, University of Newcastle \\
 Newcastle upon Tyne NE1 7RU,
 UK\cite{email}
}


\maketitle
\begin{abstract}
In this paper we   derive  exact quantum Langevin equations
for stochastic dynamics of   large-scale inflation in de~Sitter
space. These quantum Langevin equations are the equivalent
of the Wigner equation and are described by a system of stochastic
differential equations.

We   present  a formula  for the calculation of the expectation
value of a quantum operator  whose Weyl symbol is a function of
the large-scale inflation scalar field and its time derivative.
The quantum expectation value is calculated as a mathematical
expectation value over a stochastic process in an extended phase
space, where the additional coordinate plays the role of a
stochastic phase.

The unique solution is obtained for the Cauchy problem for the
Wigner equation for large-scale inflation.  The stationary solution
for the Wigner equation is found for an arbitrary potential.

It is shown that the large-scale inflation scalar field in de Sitter
space behaves as a quantum one-dimensional dissipative system,
which supports the earlier results of Graziani  and of Nakao, Nambu
and Sasaki. But the analogy with a one-dimensional model of the quantum
linearly damped anharmonic oscillator is not complete: the difference
arises from the new time dependent commutation relation for the large-scale
field and its time derivative.

It is found that, for the large-scale inflation scalar field
the large time asymptotics is equal to the `classical limit'. For the large
time
limit the  quantum Langevin equations are just the classical stochastic
Langevin equations (only the stationary state is defined by the quantum
field theory).
\end{abstract}

\section{INTRODUCTION}
\label{intro}
While the quasi-classical picture of the inflationary
universe scenario, based on a Fokker-Planck evolution
equation for the probability distribution of the inflation
field, is almost complete (see the basic papers
\cite{1,2,3,4,5,6,7,8} ),
 the essentially quantum-mechanical features of inflation
are now a subject of investigation.

We would like to note recent investigations into this
problem carried out by Graziani \cite{9,10,11,12}, Nakao, Nambu
and Sasaki \cite{13}, and Hu, Paz and Zhang \cite{14}. (The aim is
not to present the full list of papers on this subject but to mention
those which are closest to our considerations.)

In \cite{13} the dynamics of an inflationary scalar field
in a de~Sitter background is investigated on the basis of the
extended version of the stochastic approach proposed by
Starobinsky \cite{5}. In this approach,
the scalar field operator is split into
the long wavelength mode and the short wavelength mode.
This split allows the reduction of
the operator  equation for the scalar field
to Langevin equations of order $\sqrt{\hbar}$.

In the series of papers \cite{9,10,11,12} of Graziani, dealing with
quantum probability distributions and the dynamics of the
early universe, the approach is based on the Wigner function and its evolution
equation, which is the Wigner equation. The author concentrats attention
on the large-scale inflation ,where the spatial variable of
the inflation scalar field is removed by averaging over a causal
horizon volume. It is established that the
quantum description
of large-scale inflation in de~Sitter space is equal
to the quantum mechanics of a one-dimensional dissipative system.
In \cite{10} it is shown how quantum Langevin equations can
be derived to any order of $\sqrt{\hbar}$ when they correspond to
the Wigner equation expanded in powers of $\hbar$ and
truncated at some power of $\hbar$. However the Wigner equation
for large-scale inflation presented in \cite{9,10,11,12} is not
accurate and we will return to this point in our paper.

In paper \cite{14} dealing with quantum Brownian motion
Hu, Paz and Zhang have derived an exact master equation
for a quantum open system, which is an extension of
earlier results obtained by Dekker \cite{15}, and
Caldeira and Leggett \cite{16}. As was established by Vilenkin
\cite{1}, Linde \cite{3} and Starobinsky \cite{5}, the
basic stochastic equation for the large-scale quantum evolution
of inflation is similar to Brownian motion,
that is the quantum Langevin equations of order
$\sqrt{\hbar}$. So the investigation of quantum Brownian motion
is closely connected to further investigations of statistical and
quantum effects at the early universe.

In this paper we restrict our consideration of the inflation scalar field to
investigation of the dynamics of the large-scale or coarse-grained ($\geq$
causal
horizon) scalar field in a de~Sitter background. Following in the main line
Graziani
\cite{9,10,11,12} we describe the evolution of the large-scale inflation by the
Wigner equation. Then the aim of this work is to derive exact quantum Langevin
equations (to all orders of $\sqrt{\hbar}$), which describe the stochastic
dynamics of the
large-scale inflation.

Thus, our approach is opposite in some sense to the approach of Hu, Paz and
Zhang
\cite{14}. These authors move toward the possible observation of macroscopic
effects
from the search for an adequate description for statistical and quantum
effects, while our way is to start from the macro-level to obtain an equivalent
stochastic description.

\section{BASIC FORMULATIONS}
\label{basic}
The Lagrangian density of the inflation scalar field $\Phi (x,t)$ in
a de~Sitter background is
\begin{equation}
{\cal L} = -\sqrt{-g} \biggl[ \,\case 1/2 g^{\mu\nu} \partial_\mu \Phi\,
\partial_\nu \Phi + {\cal V}(\Phi )\biggr],
\label{1}
\end{equation}
where the background metric is assumed to have the form
\begin{equation}
ds^2=-dt^2+a^2(t)dx^2,
\label{2}
\end{equation}
$g$ denotes a metric, $t$ is a time, $x$ is a three-dimensional spatial
coordinate,
${\cal V}(\Phi )$ is a potential, and $a(t)$ is a scale factor (some positive
function).

Then the equation for the classical inflation scalar field $\Phi (x,t)$ is
\begin{equation}
\biggl[\frac{\partial^2}{\partial t^2}
+3\frac{\dot a (t)}{a(t)} \frac{\partial}{\partial t}
-\biggl( \frac{1}{a(t)}\biggr)^2\Delta_x\biggr] \Phi  (x,t)
+{\cal V }^{\prime}[\Phi (x,t)] =0,
\label{3}
\end{equation}
where $\Delta_x$ is the spatial Laplace operator , the dot over a symbol means
a time derivative $\dot y (t)=dy/dt$, and ${\cal V }^{\prime}[\Phi] =
d{\cal V }[\Phi]/d\Phi$.

The expansion of the universe is assumed,
\begin{equation}
\frac{\dot a (t)}{a(t)} =H(t),
\label{4}
\end{equation}
where $H(t)$ is some non-negative integrable function of time.
In other words, the scale factor is
\begin{equation}
a(t)= a(0)\exp\{\int^t_0 H(\tau)d\tau\},
\label{5}
\end{equation}
where $a(0)$ is the value of the scale factor at $t=0$, the beginning of
inflation. In this model the beginning of inflation is taken to coincide
with the origin of the universe.

Strictly speaking, in our consideration we deal with an expanding
Friedmann-Robertson-Walker (FRW) universe, `approximatly' close to a de~Sitter
universe in the sense: $H(t)\approx{\it Const}$.

{}From (1) one obtains the canonical momentum conjugate to the field
$\Phi (x,t)$ :
\begin{equation}
\Pi (x,t) =\frac{\partial{\cal L}}{\partial (\partial\Phi/ \partial t)} =a(t)^3
\frac{\partial\Phi (x,t)}{\partial t}
\label{6}
\end{equation}

and the Hamiltonian density :
\begin{eqnarray}
{\rm H} &=& \Pi\frac{\partial\Phi}{\partial t} -{\cal L}
\nonumber
\\
&=& \case1/2 a(t)^{-3} \Pi^2 +\case1/2 a(t)[\nabla_x \Phi ]^2
+a(t)^3 {\cal V}(\Phi ).
\label{7}
\end{eqnarray}

When interested in large-scale ($\geq$ causal horizon) physics,
a coarse-graining procedure is utilised and it leads to a coarse-grained or
averaged
scalar field $\Phi_X(t)$
\begin{equation}
\Phi_X(t) =\frac1V \int_{\Omega_X}\Phi (x,t)dx,
\label{8}
\end{equation}
where the index $X$ is a label referring to the center of a region $\Omega$,
over which $\Phi (x,t)$ is averaged, and $V$ is its volume. The volume of
spatial averaging is taken not smaller than a causal
horizon volume:
\begin{equation}
V\geq \case 4/3\pi\ell^3(t)
\label{9}
\end{equation}
with the causal horizon (or the `coordinate horizon' in terms of Ellis
\cite{17} ) given by
\[
\ell (t)= \int^t_0 a(\tau)^{-1}d\tau.
\]
For de~Sitter space
\begin{equation}
a(t)=\exp \{Ht\},\qquad H=\hbox{\it Const}
\label{10}
\end{equation}
then $\ell (t)=H^{-1}(1-e^{-Ht})$ and the volume of averaging is chosen to be
$V= (4/3)\pi H^{-3}$ .

For `approximatly' de~Sitter space
we will think about a volume of averaging as constant for all time. This always
can be assumed if $\ell(t)\leq {\it Const}$. The techniques we use allow
consider time dependent volume $V(t)$ along the same lines but we
shall not consider this case for the sake of brevity. (See remarks in Section~
\ref{conclusions}.)

Each large-scale region $\Omega$ (labeled by $X$) can be considered as a
separate quantum-mechanical system because each lies outside of its neighbours'
light cone: there is no exchange of information between large-scale regions.
The profit of the coarse-graining procedure is that it reduces the quantum
field problem to a quantum mechanical problem. At the same time there are still
some peculiarities, following from the field theory, which do not make the
analogy with quantum mechanics complete. We will point out these peculiarities
in this section.

After performing the coarse-graining (\ref{8}) the spatial varying term in the
Euler- Lagrange eqation (\ref{3}) can be neglected due to a smaller factor
$a^{-2}(t)V^{-2/3}$
(see \cite{18}) and the equation for the large-scale inflation field
$\Phi_X (t)$ becomes

\begin{equation}
\ddot\Phi_X (t)+3H(t)\dot\Phi_X (t)+ U ^\prime [\Phi_X (t)] =0,
\label{11}
\end{equation}
where $U(\Phi )$ is the coarse-grained potential.

The averaging of the momentum (\ref{6}) gives
\begin{equation}
\Pi_X(t) =a(t)^3 \dot\Phi_X(t).
\label{12}
\end{equation}

The Lagrangian (density) for the coarse-grained field $\Phi_X(t)$ is
\begin{equation}
{\cal L}[\Phi_X] =a(t)^3 [\case 1/2 \dot\Phi_X^2 -U(\Phi_X)].
\label{13}
\end{equation}
This gives the Euler-Lagrange equation (\ref{11}) and keeps the averaged
momentum (\ref{12}) canonically conjugate to  $\Phi_X(t)$:
\begin{equation}
\Pi_X(t) =\frac{\partial{\cal L}[\Phi_X]}{\partial \dot\Phi_X}.
\label{14}
\end{equation}

Now the Hamiltonian (density) for  the coarse-grained field $\Phi_X(t)$ is
\begin{equation}
{\cal H}(\Phi_X  ,\Pi_X ;t ) = \case1/2 a(t)^{-3} \Pi_X^2 +a(t)^3 U(\Phi_X ),
\label{15}
\end{equation}
it can be considered as the classical Hamiltonian for the large-scale inflation
field, in the sense  that the equations of motion produced by this Hamiltonian
\begin{equation}
\frac{\partial\Phi_X}{\partial t}=\frac{\partial{\cal H}}{\partial\Pi_X},
\frac{\partial\Pi_X}{\partial t}=-\frac{\partial{\cal H}}{\partial\Phi_X}
\label{16}
\end{equation}
are equivalent to the field equation (\ref{11}).

To make the step from the classical equation (\ref{11}) to the quantum equation
it is necessary to quantize (\ref{11}) taking into account (\ref{12}) and
(\ref{15}) . This leads at
the end to the quantum mechanics of a one-dimensional dissipative system in the
description on $\Phi_X $,\ $\dot\Phi_X $ variables.

One can apply the canonical procedure, which is based on quantal noise
operators and conserves the fundamental commutator for canonical position and
momentum operators in the course of time, or the influence-functional method of
Feynman and Vernon \cite{19}. Both ways give the master equation for the
`reduced' density operator $\hat\rho$ (see \cite{15,16} for techniques),
describing the time-evolution for the large-scale inflation,
\begin{equation}
\frac{\partial{\hat\rho}}{\partial t}
= \frac{V}{i\hbar} [{\hat{\cal H}},\hat\rho ] -\frac{V}\hbar
a(t)^6 D(t)[\hat\Phi_X ,[\hat\Phi_X ,\hat\rho ]].
\label{17}
\end{equation}
Here ${\hat{\cal H}}$ is an operator form for the classical Hamiltonian
(\ref{15}),
$[\,,\,]$ stands for a commutator, and the diffusion coefficient $D(t)$ is, in
general, some
non-negative function. In particular, the diffusion coefficient can be assumed
to be
$D(t)=3H(t)\cdot$~{\it Const,} where {\it Const} is determined by physical
parameters of the system
at equilibrium.

Some one who would follow, formally, the quantisation procedure in quantum
mechanics would notice immediately two points by which the master eqation
(\ref{17}) differs from the formally obtained one: firstly, the new constant
$\hbar /V$
instead of Planck's constant $\hbar$, and secondly, the factor $a(t)^6$ in the
diffusion term, which reflects the explicit time dependence of the Hamiltonian
(\ref{15}). It is time to discuss how these peculiarities come from the quantum
field theory.

The origin of the scaling
\begin{equation}
\hbar \to \hbar /V
\label{18}
\end{equation}
in equation (\ref{17}), where $V$ is a volume of averaging (\ref{9}), is that
the
fundamental commutator for canonical position and momentum operators in the
relativistic
quantum field theory
\[
[\hat\Phi  (x,t),\hat\Pi (y,t)] =i\hbar\,\delta (x-y)
\]
is transformed by coarse-graining procedure (\ref{8}) to
\begin{equation}
[{\hat\Phi }_X (t),{\hat\Pi}_X(t)]  =  i\hbar /V,
\label{19}
\end{equation}
\[
[{\hat\Phi }_X (t),{\hat\Pi}_Y (t)]  =   0,\qquad  X\neq Y.
\]

An other way to derive the master equation (\ref{17}) would be to use the
following momentum and Hamiltonian:
\begin{equation}
\Pi_\Omega (t) = V\Pi_X(t),
\label{20}
\end{equation}
\[
{\cal H}_\Omega = V{\cal H}(\Phi_X, \Pi_X; t).
\]
For momentum $\Pi_\Omega$ the fundamental commutator is
\begin{equation}
[{\hat\Phi }_X (t),{\hat\Pi}_\Omega(t)]  =  i\hbar.
\label{21}
\end{equation}
The description in terms of (\ref{20})- (\ref{21}) is natural for the
large-scale region $\Omega$. Returning back to the spatially dependent field
$\Phi (x,t)$, the Hamiltonian for $\Omega$ is found to be 
\begin{equation}
{\cal H}_\Omega = \int_\Omega {\rm H}(x,t)dx,
\label{22}
\end{equation}
or, in view of the Hamiltonian (density) (\ref{15}),
\begin{equation}
{\cal H}_\Omega = V{\cal H}(\Phi_X, \Pi_X; t)
\label{23}
\end{equation}
(in (\ref{23}) the spatially varying term of the original field
$[\nabla_x\Phi]^2$ was neglected ).

Now let us discuss the diffusion term in the master equation (\ref{17}).
It should be mentioned that the general form of the diffusion term in the
master
equation is
\[
-\frac{V^2}{\hbar^2} C(t)[\hat\Phi ,[\hat\Phi ,\hat\rho\, ]],
\]
where $C(t)$ is some time-dependent coefficient. This coefficient is defined
with respect to a suitably chosen vacuum state in the field theory. For
de~Sitter space it is the so-called Bunch- Davies vacuum \cite{20}-\cite{21}.
In terms
of (\ref{17}) this means that it is determined by the physical parameters of
the system at equilibrium.

If in accordance with the field theory (of inflation) we assume that the energy
density of the equilibrium state is an invariant in de~Sitter space then
\begin{equation}
\langle{\hat{\cal H}}_\Omega (t)\rangle / a(t)^3V \,= \langle{\hat{\cal
H}}({\hat\Phi}_X,{\hat\Pi}_X;t)\rangle/a(t)^3 = {\it Const},
\label{24}
\end{equation}
where, to obtain the energy density, the Hamiltonian ${\cal H}_\Omega$ is
divided by the proper volume $ a(t)^3V$. The expectation values,
$\langle\,\,\rangle$    , of the operators are taken on the stationary solution
of the master equation. For `approximatly' de~Sitter space, $\dot H(t)$ is
neglected so the condition (\ref{24}) can be assumed.

To fulfil (\ref{24}) we have obtained the factor $a(t)^6$ in the diffusion term
of (\ref{17}). Taking into account that the constant in (\ref{24}) is
proportional to $(\hbar/V)$, in (\ref{17}) we can show explicitly the
dependence
\[
C(t)= {\hbar \over V} a(t)^6 D(t),
\]
which is valid for large-scale inflation.
We will use master equation (\ref{17})  as a starting point for investigation
of the stochastic dynamics of the large-scale inflation.

{}From this point onwards in the paper the large-scale scalar field is refered
to as $\Phi (t)$, omitting the index $X$.

\section{THE WIGNER EQUATION FOR LARGE-SCALE INFLATION}
\label{wigner}
The Wigner function $W(q,p;t)$ \cite{22,23} is a function on the classical
phase space
and describes the distribution of position and momentum. The Wigner function is
not a
probability distribution since it can assume negative values; $W(q,p;t)$ is a
real
function.

As a density matrix the Wigner function contains all of the information
corresponding to the quantum state. The expectation value for an arbitrary
operator ${\hat A}({\hat\Phi},{\hat{\dot\Phi}})$
can be calculated by the formula
\begin{equation}
\langle {\hat A}\rangle =\int dq\, dp\, A(q,p)W(q,p;t),
\label{25}
\end{equation}
where $A(q,p)$ is a Weyl symbol for the operator ${\hat A} ({\hat\Phi}
,{\hat{\dot\Phi}})$.

Let us derive the time-evolution equation for the Wigner function for
large-scale inflation, that is equivalent to the master equation (\ref{17}).

For the Wigner operator $\hat W$ the equation has the same form as for the
density
operator $\hat\rho$ (\ref{17}) (because coefficients
$a(t)$ and $D(t)$ depend only on $t$)
\begin{equation}
{\partial {\hat W} \over \partial t} = {V\over i\hbar}
[ {\hat{\cal H}}, {\hat W} ] - {V\over \hbar} \
a(t)^6 D(t) [{\hat q},[{\hat q},{\hat W}]],
\label{26}
\end{equation}
where Weyl symbols for ${\hat W}$ and ${\hat{\cal H}}$ are
\begin{eqnarray}
&{\hat W} \leftrightarrow W(q,\Pi;t),
\nonumber\\
&{\hat {\cal H}} \leftrightarrow {\cal H}(q,\Pi) = \case1/2 a(t)^{-3} \Pi^2 +
a(t)^3 V(q).
\label{27}
\end{eqnarray}
Here we assume the following correspondence between phase space variables
$(q,\Pi)$ or $(q,p)$ and variables of the large-scale inflation scalar field
$(\Phi, {\dot \Phi})$:
\begin{equation}
\Phi =q, \qquad {\dot \Phi} =p,\qquad a(t)^3{\dot \Phi} =\Pi.
\label{28}
\end{equation}

To obtain an equation for the Wigner function $W(q,\Pi;t)$
from its operator form (\ref{26}) we will use the formula
for the composition of operators in the Weyl formalism.
Due to this formula \cite{24}
and the commutator relation (\ref{19})
a Weyl symbol $A(q,\Pi)$ for the
composition of two operators ${\hat A} = {\hat B} {\hat C}$ is defined via the
correspondence
\begin{eqnarray}
A(q,\Pi) &=& B\biggl( \qtwo +\frac{i\hbar}{2V}^I \frac{\partial}{\partial \Pi},
\pitwo -\frac{i\hbar}{2V}^I \frac{\partial}{\partial q}\biggr) C(q,\Pi)
\nonumber\\
&=& C\biggl( \qtwo -\frac{i\hbar}{2V}^I \frac{\partial}{\partial \Pi},
\pitwo +\frac{i\hbar}{2V}^I \frac{\partial}{\partial q}\biggr) B(q,\Pi)
\label{29}
\end{eqnarray}
where $B(q,\Pi )$ and $C(q,\Pi )$  are Weyl symbols for the operators $\hat B$
and $\hat C$, and the numbers I,II over operators show the order in which the
operators act.

Using (\ref{27}),(\ref{29}) one can obtain the following correspondence
between operators and their Weyl symbols:
\begin{eqnarray}
[\hat{\cal H},\hat W ] &\leftrightarrow& -{i\hbar \over V}\, a(t)^{-3} \Pi
{\partial \over \partial q} W(q,\Pi;t)  \nonumber\\
& &+ a(t)^3 \biggl[ U\biggl(q+{i\hbar\over 2V}
{\partial \over \partial \Pi} \biggr) -U\biggl(q-{i\hbar\over 2V}
{\partial \over \partial \Pi} \biggr) \biggr]\, W(q,\Pi;t),
\label{30}
\end{eqnarray}
\begin{equation}
[{\hat q},[{\hat q},{\hat W}]] \leftrightarrow -{\hbar^2\over V^2}
{\partial^2 \over \partial\Pi^2} W(q,\Pi;t).
\label{31}
\end{equation}

Combining relations (\ref{30})--(\ref{31}) in accordance with equation
(\ref{26}) one has the time evolution equation for the Wigner function
$W(q,\Pi;t)$
\begin{eqnarray}
{\partial W(q,\Pi;t)\over \partial t}  &=& -a(t)^{-3} \Pi
{\partial  W \over \partial q} +{\hbar \over V} a(t)^6 D(t)
{\partial^2  W \over \partial \Pi^2}
\nonumber\\
& &+{V\over i\hbar} a(t)^3 \biggl[U\biggl(q+{i\hbar\over 2V}
 {\partial    \over \partial\Pi}\biggr)
-U\biggl(q-{i\hbar\over 2V} {\partial    \over \partial\Pi}\biggr)\biggr]
W(q,\Pi;t)
\label{32}
\end{eqnarray}

Equation (\ref{32}) is called the Wigner equation. (Note that the
commonly used way to derive the Wigner equation is to apply
the coordinate representation for the Wigner function
\begin{equation}
W(q,\Pi;t) = {1\over 2\pi} \int\limits_{-\infty}^\infty e^{i\Pi x}
\biggl< q-{\hbar \over 2V} x \biggm|{\hat \rho}\biggm|  q+{\hbar \over 2V}
x\,\biggr> \, dx
\label{33}
\end{equation}
to the master equation (\ref{17}). Actually it is a longer way
because it deals with multiple integrations and integrations by parts
while the Weyl formalism gives an answer in a plain
algebraic way.)

Because we are interested in the distribution of $\Phi$ and ${\dot \Phi}$
let us make the transition from $W(q,\Pi;t)$ to $W(q,p;t)$ by
relations (\ref{28}) and equality
\begin{equation}
W(q,p;t) =a(t)^3 W(q,\Pi;t)
\label{34}
\end{equation}
which follows from (\ref{25}). For the Wigner function $W(q,p;t)$ one has
the time evolution equation
\begin{eqnarray}
\frac{\partial W(q,p;t)}{\partial t}
&=& -p\frac{\partial W}{\partial q} +3H(t)\frac\partial{\partial p} (pW)
+\frac\hbar{V}  D(t)\frac{\partial^2W}{\partial p^2}
\nonumber\\
& & +\frac{V}{i\hbar} a(t)^3
\biggl[ U\biggl(q+\frac{i\hbar}{2V} a(t)^{-3}\frac\partial{\partial p}\biggr)
-U\biggl(q-\frac{i\hbar}{2V} a(t)^{-3}\frac\partial{\partial p}\biggr) \biggr]
W(q,p;t),
\label{35}
\end{eqnarray}
where $H(t)$ is defined by (\ref{4}).

The Wigner equation (\ref{35}) can be rewritten in the following equivalent
form:
\begin{eqnarray}
\frac{\partial W(q,p;t)}{\partial t}
&=& -p\frac{\partial W}{\partial q} +3H(t)\frac\partial{\partial p} (pW)
+\frac\hbar{V} D(t)\frac{\partial^2W}{\partial p^2}
\nonumber\\
& &+ \frac{V}{i\hbar} a(t)^3
\int^\infty_{-\infty} du\,W(q,p-u;t)\cdot {\cal I}
(q,u;t),
\label{36}
\end{eqnarray}
where
\begin{equation}
{\cal I}(q,u;t) =\frac1{2\pi} \int^\infty_{-\infty} dy\,e^{-iuy}
\biggl[ U\biggl(q+\frac\hbar{2V} a(t)^{-3}y\biggr)
-U\biggl(q-\frac\hbar{2V} a(t)^{-3}y\biggr)\biggr].
\label{37}
\end{equation}

Equation (\ref{35}), or equation (\ref{36}), is the complete Wigner equation
(to all
orders of $\hbar$) for the large-scale inflation in de~Sitter space. It
describes the
time evolution for the distribution of $\Phi$ and $\dot\Phi$ in a sense of
equality (\ref{25}). The expansion of the Universe, which is described by the
scale factor $a(t)$ and equality (\ref{4}), gives the origin for a dissipation
term, with the disspation coefficient $3H(t)$, in the Wigner equation. In this
sense large-scale inflation can be considered as a quantum one-dimensional
dissipative system, which supports Graziani's statement \cite{9}.
At the same time the expansion of the Universe influences the potential
terms of equation (\ref{35}) (or (\ref{36})), which was missed in
\cite{9,10,11,12}.

In \cite{15,14} the Wigner equation for a quantum disspative system
was derived for a harmonic oscillator's potential $U(q)=\omega^2q^2/2$.
For an arbitrary potential $U(q)$ a truncated Wigner equation, or a
Kramers--Moyal equation, is commonly used in the literature instead of the
complete Wigner equation (see for example \cite{16}). In our case (\ref{35}) it
would be
\begin{eqnarray}
\frac{\partial W(q,p;t)}{\partial t}
&=& -p\frac{\partial W}{\partial q} +3H(t)\frac\partial{\partial p} (pW)
+\frac\hbar{V} D(t)\frac{\partial^2W}{\partial p^2}
\nonumber\\
& &+ \frac{\partial U(q)}{\partial q}\cdot \frac{\partial W}{\partial p}
+O(\hbar^2),
\label{38}
\end{eqnarray}
where $O(\hbar^2)$ is a value of order $\hbar ^2$.

The potential term in equation (\ref{35}) can be expanded in powers of $\hbar$
by
Taylor's series, which leads to equation
\begin{eqnarray}
\frac{\partial W(q,p;t)}{\partial t}
&=& -p\frac{\partial W}{\partial q} +3H(t)\frac\partial{\partial p} (pW)
+\frac\hbar{V} D(t)\frac{\partial^2W}{\partial p^2}
\nonumber\\
& &+ \sum^\infty_{k=1,3,5,\dots} \frac1{k!}
\biggl(\frac{i\hbar}{2V} a(t)^{-3}\biggr)^{k-1}
\frac{\partial^k U(q)}{\partial q^k}
\frac{\partial^kW}{\partial p^k} .
\label{39}
\end{eqnarray}

In the series of papers \cite{9,10,11,12} the expansion in powers of $\hbar$ of
the Wigner equation was used to show how to improve the accuracy of (\ref{38}).

The potential term in Wigner equation (\ref{35}) (or in the equivalent
representations (\ref{36})--(\ref{37}) and (\ref{39})) on a macro-level
contains information about
quantum noise on the micro-level, when
\begin{equation}
\frac{\partial^kU(q)}{\partial q^k} \not\equiv 0\qquad\mbox{for}\ k\geq 3,
\label{40}
\end{equation}
This quantum noise is non-Gaussian noise. Condition (\ref{40}) is always
fulfilled when a potential deviates from a harmonic oscillator's potential. The
diffusion term, containing $\partial^2W/\partial p^2$, in the Wigner equation
also represents quantum noise (Gaussian noise) on a macro-level.
Some authors consider a possibility that the Wigner equation may contain second
order derivatives $\partial^2W/\partial q^2$,
$\partial^2W/\partial q\partial p$, which are also of quantum origin
\cite{15,14}.
(These terms would correspond to additional terms
$[\hat\Pi ,[\hat\Pi ,\hat\rho\,]]$,
$[\hat\Phi ,[\hat\Pi ,\hat\rho\,]]$,
$[\hat\Pi ,[\hat\Phi ,\hat\rho\,]]$ in the master equation (17).) These terms
have no physical
sense because they lead to  wrong Langevin equations:
$dq_t/dt\not= p_t$ or $d\Phi /dt\not= \dot\Phi$.

In this paper we will work with the complete Wigner equation
(\ref{36})--(\ref{37})
for large-scale inflation to derive equations for the corresponding stochastic
dynamics.
Let us supply the evolution equation (\ref{36}) with the initial condition
\begin{equation}
W(q,p;0)=W_0(q,p),
\label{41}
\end{equation}
where the initial Wigner function is chosen to satisfy the following
properties:
\begin{eqnarray}
\mbox{a)}\quad&W_0(q,p)\hbox{ is a real function,}
\label{42}
\\
\mbox{b)}\quad&\int W_0(q,p)dq\,dp =1,
\label{43}
\\
\mbox{c)}\quad&\int [W_0(q,p)]^2dq\,dp\leq \frac{V}{2\pi\hbar}
a(0)^3,
\label{44}
\end{eqnarray}

\noindent which follows from general properties for the Wigner function
(see \cite{23}).

\section{QUANTUM LANGEVIN EQUATIONS}
\label{langevin}
In this section we deduce quantum Langevin equations which are equal to the
complete
Wigner equation (\ref{36})--(\ref{37}), but we start with some assumption about
the potential $U(q)$.
\begin{equation}
U(q)=\frac{\omega^2}{2} q^2 +\int\exp \{ -iqp\} \,\mu (dp),
\label{45}
\end{equation}
where $\mu (p)$ is a bounded measure such that
\begin{eqnarray}
\int \mu (dp) &\leq& \mbox{\it Const},
\nonumber\\
\int p^2 \mu (dp) &\leq& \mbox{\it Const}.
\label{46}
\end{eqnarray}
The first term in the right hand side of equation (\ref{45}) is just the
harmonic
oscillator's potential while the second term can be considered as a deviation
from it. The potential is assumed to be real.

To include more model potentials used in the theory of inflation $\omega^2$ is
allowed to be either positive or negative, or to be zero. A parabolic potential
connected with `chaotic inflation' is included in (\ref{45}) with
$\mu(dp)\equiv 0$.
A double-well potential, representable in the form (\ref{45}) is, for example
\begin{equation}
U(q)=\omega^2q^2/2 + K \cos (kq)\cdot
I_{[-(3\pi /2k),(3\pi /2k)]}\,(q),
\label{47}
\end{equation}
where $\omega$,$K$,$k$ are some parameters (real values) and $I_A(q)$ is the
identificator of a set $A$:
\[
I_A(q)=\cases{1, &if $q\in A$\cr 0, &if $q\not\in A$.\cr}
\]

Form (\ref{45}) rules out model potentials which are polynomials higher than
second
order in $q$. Remember here that model potentials of polynomial type have
appeared in
the theory of inflation through expansion of $U(q)$ in powers of $q$ near $q=0$
(see,
for example, \cite{5}). Thus, almost all physical potentials for the inflation
field
$\Phi (t)$ can be represented as (\ref{45}).

For potential (\ref{45}) the Wigner equation becomes
\begin{eqnarray}
\frac{\partial W(q,p;t)}{\partial t}
&=& -p \frac{\partial W}{\partial q} +\omega^2 q\frac{\partial W}{\partial p}
+3H(t)\frac{\partial}{\partial p} (pW)
+\frac\hbar{V} D(t)\frac{\partial^2W}{\partial p^2}
\nonumber\\
& &-\frac{iV}\hbar a(t)^3 \int d\mu (p')\exp \{ -iqp'\}
\biggl[W\biggl( q,p+\frac\hbar{2V} a(t)^{-3}p';t\biggr)
\nonumber\\
& &-W\biggl(q,p-\frac\hbar{2V} a(t)^{-3}p';t\biggr)\biggr],
\label{48}
\end{eqnarray}
$q,p,p'\in{\bf R}^1$,\ $t\in [0,T]$, supplied by the initial condition
\begin{equation}
W(q,p;0)=W_0(q,p),
\label{49}
\end{equation}
such that properties (\ref{42})--(\ref{44}) are satisfied.

In order to solve this equation and to deduce stochastic equations, which
describe a
path in a phase space $(q,p)\equiv (\Phi ,\dot\Phi )$, we should reduce it to
the form
of a forward Kolmogorov equation \cite{25}( p.102).
Such an approach was first proposed by Chebotarov and Maslov \cite{26} and
developed much further by Combe, Guerra et al. \cite{27}.

To reduce equation (\ref{48}) to a forward Kolmogorov equation we need to
transform the
last term in the right hand side of it to a standard form. It can be done by
introduction of a new function
\begin{equation}
f(q,p,\theta ;t) =W(q,p;t) \exp \biggl\{ -\frac{2V}\hbar A(t) +i\theta
\biggr\},\quad
\theta\in{\bf R}^1 /\mbox{mod}\,2\pi ,
\label{50}
\end{equation}
where $\theta$ is a new variable, and function $A(t)$ will be defined later.

Inserting (\ref{50}) into (\ref{48}) one has
\begin{eqnarray}
\frac{\partial f(q,p,\theta ;t)}{\partial t}
&=& -p \frac{\partial f}{\partial q} +\omega^2 q\frac{\partial f}{\partial p}
+3H(t)\frac{\partial}{\partial p} (pf)
 +\frac\hbar{V} D(t)\frac{\partial^2f}{\partial p^2}
\nonumber\\
& &+\frac{V}\hbar a(t)^3\int^\infty_{-\infty} du\int \mu (dp')
\biggl[f\biggl(q,p+\frac\hbar{2V} a(t)^{-3}p'u, \theta +\frac\pi{2}
u-qp';t\biggr)
\nonumber\\
& &-f(q,p,\theta ;t)\biggr]\cdot [\delta (u+1)+\delta (u-1)]
\nonumber\\
& &+\frac{2V}\hbar a(t)^3 f(q,p,\theta ;t)\cdot \int \mu (dp')
-\frac{2V}\hbar \frac{dA(t)}{dt}  f(q,p,\theta ;t),
\label{51}
\end{eqnarray}
where $\delta (u)$ is a Dirac delta-function.

Let us put
\begin{equation}
A(t) =\int^t_0 a(\tau )^3 d\tau\cdot\int \mu (dp).
\label{52}
\end{equation}
Note that for scale factor (\ref{10})
\begin{equation} A(t) =\int \mu (dp)\cdot (e^{3Ht}-1)/3H.
\label{53}
\end{equation}
Let $m_\hbar (dpdu;t)$ be the following measure:
\begin{equation}
m_\hbar (dpdu;t) =\frac{V}\hbar a(t)^3 \mu (dp)[\delta (u+1)+\delta (u-1)] du.
\label{54}
\end{equation}
The measure $m_\hbar (dp\,du;t)$ is a bounded measure on
${\bf R}^1\otimes (-1)\otimes (+1)$ multiplied by time-dependent function
$a(t)^3$. For
an arbitrary function $\psi (p,u)$ an integral over this measure is
\begin{equation}
\int\psi (p,u)m_\hbar (dp\,du;t)
=\frac{V}\hbar a(t)^3 \int [\psi (p,+1)+\psi (p,-1)] \mu (dp).
\label{55}
\end{equation}

Now, with (\ref{52}) and (\ref{54}), equation (\ref{51}) becomes
\begin{eqnarray}
\frac{\partial f(q,p,\theta ;t)}{\partial t}
&=& -p\, \frac{\partial f}{\partial q} +\omega^2 q\,\frac{\partial f}{\partial
p}+3H(t)\frac{\partial}{\partial p} (pf)
+\frac\hbar{V} D(t)\,\frac{\partial^2f}{\partial p^2}
\nonumber\\
& &+\int \biggl[f \biggl(q,p+\frac\hbar{2V} a(t)^{-3}p'u, \theta -\frac\pi{2}
u-qp';t\biggr)
\nonumber\\
& &-f(q,p,\theta ;t)\biggr]m_\hbar (dp'du;t),
\label{56}
\end{eqnarray}
which is the forward Kolmogorov equation.

If equation (\ref{56}) is formally supplied by the initial condition
\begin{equation}
f(q,p,\theta ;t) =\delta (q-q_0)\,\delta (p-p_0)\,\delta (\theta -\theta_0),
\label{57}
\end{equation}
then, according to the theory of stochastic differential equations (SDE), there
exists a
three-dimensional stochastic process
\begin{equation}
\xi_t \equiv (\Phi_t,\dot\Phi_t ,\Theta_t)
\label{58}
\end{equation}
for which the function $f(q,p,\theta ;t)$ is a probability distribution. This
means that
for an arbitrary function
$h(q,p,\theta )$, which is continuous and periodic with the period $2\pi$ on
variable
$\theta$ and for fixed $\theta$ belongs to the class
${\bf C}^{1,2}_b ({\bf R}^1\otimes {\bf R}^1$),
\begin{equation}
\int h(q,p,\theta )f(q,p,\theta ;t) dq\,dp\,d\theta
={\bf E}h(\xi_t),
\label{59}
\end{equation}
where the symbol ${\bf E}$ denotes mathematical expectation and
\begin{equation}
\xi_{t=0}=(q_0,p_0,\theta_0)
\label{60}
\end{equation}
(see equality (\ref{57})).

Using the generalized Ito formula for stochastic differentials (see
\cite{25},p.270, formula (13)) one deduces an SDE for the stochastic process
$\xi_t$
(\ref{58})
\begin{eqnarray}
d\Phi_t &=& \dot\Phi_t dt,
\label{61}
\\
d\dot\Phi_t &=&
-[3H(t)\dot\Phi_t +\omega^2\Phi_t]dt
+\biggl[ \frac{2\hbar}V D(t)\biggr]^{1/2}dw_t
\nonumber\\
& &-\frac\hbar{2V} a(t)^{-3} \int pu\nu_\hbar (dp\,du;dt),
\label{62}
\\
d\Theta_t &=& \int \biggl[\frac\pi{2} u+\Phi_t p\biggr] \nu_\hbar (dp\,du;dt).
\label{63}
\end{eqnarray}
Here $w_t$ is a one-dimensional Wiener process (or Brownian motion) and
$\nu_\hbar (dp\,du;dt)$ is a Poisson measure on
${\bf R}'\otimes (-1)\otimes (+1)\otimes [0,T]$ non-homogeneous with respect
to translation on $[0,T]$ such that
\begin{equation}
{\bf E}[\nu_\hbar (dp\,du;dt)]= m_\hbar (dp\,du;t) dt.
\label{64}
\end{equation}
$\Theta_t$ is an additional stochastic variable, which can be
interpreted as a stochastic phase (see equation (\ref{50})).
It appears only due to deviation of the potential (\ref{45}) from the harmonic
oscillator's potential.

In order for the stochastic process (\ref{61})--(\ref{63}) to have a unique
solution on a time interval $[0,T]$, right continuous with probability 1, it is
enough to have conditions (\ref{46}), and to assume that functions $a(t)$,
$H(t)$, $a(t)^{-3}$, $D(t)$ and their first derivatives are continuous
functions.

Let us denote the stochastic process (\ref{61})--(\ref{63}) with initial
condition (\ref{57}) by
\begin{equation}
\xi_t(q_0,p_0,\theta_0)
=(\Phi_t(q_0,p_0,\theta_0),\dot\Phi_t(q_0,p_0,\theta_0),
\Theta_t(q_0,p_0,\theta_0)).
\label{65}
\end{equation}

If the initial condition for equation (\ref{56})
\begin{equation}
f(q,p,\theta ;0)=f_0(q,p,\theta )
\label{66}
\end{equation}
belongs to the class of generalized functions then instead of (\ref{59}) one
has
\begin{eqnarray}
\int\! &h&(q,p,\theta )f(q,p,\theta ;t)dq\, dp\, d\theta =
\nonumber\\
&=& \int {\bf E}h(\Phi_t(q,p,\theta),\dot\Phi_t(q,p,\theta),
\Theta_t(q,p,\theta)) f_0(q,p,\theta)dq\, dp\, d\theta .
\label{67}
\end{eqnarray}

Now one can readily find the correspondence between an integral over the Wigner
function $W(q,p;t)$, governed by equation (\ref{48}), and the stochastic
process
 (\ref{61})--(\ref{63}). Let us assume that
\[
f_0(q,p,\theta) =W_0(q,p)
\]
and let the function $h(q,p,\theta)$ in (\ref{67}) be of the form
\[
h(q,p,\theta) = h(q,p)\exp\{-i\theta\}.
\]
Then from (\ref{67}) and (\ref{50}) one finds
\begin{eqnarray}
\int \! &h&(q,p )W (q,p;t)dq\, dp=
\nonumber\\
&=&\exp \biggl\{\frac{2V}\hbar A(t)\biggr\} \int\!{\bf E}
\{ h(\Phi_t(q,p,0),\dot\Phi_t(q,p,0))
\exp [-i\Theta_t(q,p,0) ]\}  W_0(q,p)dq\, dp,
\label{68}
\end{eqnarray}
where $\int d\theta =2\pi$ was used and the function $A(t)$ is defined by
(\ref{52}).

Formula (\ref{68})
gives one the expectation value of a quantum operator
${\hat A}(\hat\Phi ,\hat{\dot\Phi})$ with its Weyl symbol $h(q,p)$
(compare (\ref{68}) with (\ref{25})).

Equations (\ref{61})--(\ref{63}) are exact quantum Langevin equations for the
large-scale inflation in de~Sitter space associated with the master equation
(\ref{17}).

In \cite{10} an attempt was made to derive the quantum Langevin equations for
potentials
with a polynomial growth higher than second order in $q$. As  was mentioned,
such potentials are excluded in our consideration, which is restricted by
(\ref{45}). In \cite{10} the deduction of the quantum Langevin equations is
based on a general expansion in powers of $\hbar$ of the Wigner equation, as in
our representation (\ref{39}). From \cite{10} it follows that for polynomial
potentials, of order higher than second, the quantum Langevin equations can be
derived exactly only to order $\sqrt{\hbar /V}\,$ (corresponding to the
truncated Wigner equation of order $\hbar /V$). Already the first correction of
order $\hbar /V$ to the quantum Langevin equations cannot be calculated
precisely (there is no explicit representation for the
noise terms in the quantum Langevin equations). The origin of this problem is
related to the term $\partial^3W/\partial p^3$ in  the truncated Wigner
equation of order $(\hbar /V)^2$.

Let us rewrite the quantum Langevin equations (\ref{61})--(\ref{63})
in an integral form to show explicitly dependence on initial data $(q,p,0)$:
\begin{eqnarray}
\Phi_t(q,p,0) &=&
q+\int^t_0\dot\Phi_\tau (q,p,0)  d\tau ,
\label{69}
\\
\dot\Phi_t(q,p,0) &=&
p-\int^t_0
[3H(\tau )\dot\Phi_\tau (q,p,0) +\omega^2\Phi_\tau(q,p,0) ]d\tau
+\biggl[\frac{2\hbar}V\biggr]^{1/2}\int^t_0 D(\tau )^{1/2}dw_\tau
\nonumber\\
& &-\frac\hbar{2V}\int^t_0 a(\tau )^{-3}p'u\nu_\hbar (dp'du;d\tau ),
\label{70}
\\
\Theta_t(q,p,0) &=& \biggl\{ \int^t_0\int
\biggl[\frac\pi{2} u+\Phi_\tau (q,p,0)p'\biggr]\nu_\hbar (dp'du;d\tau )\biggr\}
/\mbox{mod}\, 2\pi .
\label{71}
\end{eqnarray}

Equations (\ref{69})--(\ref{71}) describe a stochastic path
$\Phi_t$,$\dot\Phi_t$ in phase space $(\Phi ,\dot\Phi)$ starting at a point
$(q,p)$ when
$t=0$. The stochastic phase $\Theta_t$ plays role in the final formula
(\ref{68}) for
the expectation value of a quantum operator and can be interpreted as a
contribution of
the stochastic path $\Phi_t(q,p)$,$\dot\Phi (q,p)$.

Equation (\ref{68}) together with equations (\ref{61})--(\ref{63}) can
be used for numerical simulations to calculate the expectation value for
quantum operator
$\hat A (\hat\Phi ,\hat{\dot\Phi})$,
which is much simpler than solving the Wigner equation because SDE
 (\ref{61})--(\ref{63}) are of first order.

\section{SOLUTION OF THE WIGNER EQUATION}
\label{solution}
The aim of this section is to show how the Wigner function itself can be
expressed by
the expectation value with respect to a stochastic process in the extended
phase space
$(q,p,\theta )$. In \cite{27} such an expression is found for the case
\begin{equation}
\frac{\partial\hat W}{\partial t}=\frac1{i\hbar} [\hat H,\hat W],\quad
H(q,p)=\frac{p^2}{2m} +\case1/2 m\omega^2 q^2 +U(q,p)
\label{72}
\end{equation}
while in the case under consideration (\ref{26}),(\ref{15}) the diffusion term
and the expansion of the Universe play an essential role.

To find a solution for the Wigner equation (\ref{48}) one should transform it
to a backward Kolmogorov equation \cite{25}( p.300). This can be done by
introduction of the following function:
\begin{equation}
f(q,p,\theta ;t) =W(q,p;t)\exp
\biggl\{ -3\int^t_0 H(\tau )d\tau -\frac{2V}\hbar A(t) +i\theta \biggr\},
\label{73}
\end{equation}
where $\theta\in {\bf R}^1/\mbox{mod}\,2\pi$, and $A(t)$ is the same function
(\ref{52}) as in Section~\ref{langevin}.

The time evolution of $f$ is governed now by the equation
\begin{eqnarray}
\frac{\partial f(q,p,\theta ;t)}{\partial t}
&=&-p \frac{\partial f}{\partial q} +[\omega^2q +3H(t)p]\frac{\partial
f}{\partial p}
+\frac\hbar{V} D(t)\frac{\partial^2 f}{\partial p^2}
\nonumber\\
& &+\int \biggl[f\biggl(q,p+\frac\hbar{2V} a(t)^{-3} p'u,\theta -\frac\pi{2}
u-qp';t\biggr)
\nonumber\\
& &-f(q,p,\theta ;t)\biggr] m_\hbar (dp'du;t),
\label{74}
\end{eqnarray}
which would be a backward Kolmogorov equation for the backward time $t'$, by
setting
$t=T-t'$, where $0\leq t\leq T$. The measure
$m_\hbar (dp\, du;t)$ is defined by (\ref{54}).

If equation (\ref{74}) is supplied by an initial condition
\begin{equation}
f(q,p,\theta ;0) =f_0(q,p,\theta ),
\label{75}
\end{equation}
where $f_0$ is a continuous function, then the solution for
(\ref{74}) can be represented as
\begin{equation}
f(q,p,\theta ;t) ={\bf E}f_0 (Q_T(q,p,\theta ;T-t),
P_T(q,p,\theta ;T-t),\Theta_T(q,p,\theta ;T-t)),
\quad
0\leq t\leq T,
\label{76}
\end{equation}
where the stochastic process can be found by applying the generalized Ito
formula for stochastic differentials \cite{25} (p.270),
\begin{eqnarray}
Q_s(q,p,\theta ;T-t) &=&
q-\int^s_{T-t} P_\tau (q,p,\theta ;T-t) d\tau ,
\label{77}
\\
P_s(q,p,\theta ;T-t) &=&
p+\int^s_{T-t} [3H(\tau )P_\tau (q,p,\theta ;T-t)
+\omega^2Q_\tau (q,p,\theta ;T-t)] d\tau
\nonumber\\
& &+\biggl[\frac{2\hbar}V \biggr]^{1/2}
\int^s_{T-t} D(\tau )^{1/2}dw_\tau
\nonumber\\
& &+\frac\hbar{2V} \int^s_{T-t}\int a(\tau )^{-3}p'u
\nu_\hbar (dp'du ;d\tau ),
\label{78}
\\
\Theta_s(q,p,\theta ;T-t) &=&
\biggl\{ \theta -\int^s_{T-t}\int \biggl[\frac\pi{2} u
+Q_\tau (q,p,\theta ;T-t) p'\biggr]
\nu_\hbar (dp'du ;d\tau ) \biggr\}/\hbox{mod}\, 2\pi,
\label{79}
\end{eqnarray}
where $0\leq T-t\leq s\leq T$ and $\nu_\hbar (dp'du ;d\tau )$ is a Poisson
measure on ${\bf R}^1\otimes (-1)\otimes (+1)\otimes [0,T]$ with the intensity
defined by (\ref{64}).

Returning to the Wigner function one has
\begin{eqnarray}
W(q,p;t) &=& \exp \biggl\{ 3\int^t_0 H(\tau )d\tau
+\frac{2V}\hbar  A(t)\biggr\} \cdot
\nonumber\\
& &\cdot {\bf E} \{ W_0 (Q_T(q,p,0 ;T-t), P_T(q,p,0 ;T-t))
\exp [i\Theta_T(q,p,0 ;T-t)]\} ,
\label{80}
\end{eqnarray}
where the stochastic process is defined by equations (\ref{77})--(\ref{79})
with $\theta =0$.

Conditions for existence and uniqueness of the solution for SDE
(\ref{77})--(\ref{79})
are the same as for SDE (\ref{69})--(\ref{71}).

Let us assume additionally that
\begin{equation}
\int p^k \mu (dp)\leq\mbox{\it Const},\qquad k=3,4,
\label{81}
\end{equation}
and that the initial Wigner function $W_0(q,p)$ is twice continuously
differentiable in
$p$ and once in $x$, and that its first and second order partials are bounded.
Then the Wigner function (\ref{80}) is twice continuously differentiable in
$p$ and once in $x$, differentiable in $t$ and is the unique solution for the
Cauchy problem (\ref{74})--(\ref{75}).

Let us consider how properties (\ref{42})--(\ref{44}) for the initial Wigner
function are preserved in the course of time.

a) The reality of the Wigner function is preserved.
To prove this proposition it is necessary to use the decomposition of the
Wigner function into a difference of two positive functions (see \cite{27}
formula (4.4)). The group of transformation (\ref{80}) preserves this
decomposition.

b) The normalization for the Wigner function
\[
\int W(q,p;t)dq\,dp =1,\qquad t\geq 0,
\] is fulfilled due to the equality
\[
\frac{d}{dt} \int W(q,p;t)dq\, dp =0
\]
which follows directly from the Wigner equation.

c) The presence of the dissipation and diffusion terms in the Wigner equation
destroys the inequality (\ref{44}). It now becomes
\begin{equation}
\int W^2(q,p;t)dq\,dp \leq \int W_0^2 (q,p)dq\,dp\cdot
\exp \biggl\{ 3\int^t_0 H(\tau )d\tau \biggr\}
\leq \frac{V}{2\pi\hbar} a(t)^3.
\label{82}
\end{equation}

To derive (\ref{82}) it is necessary to take a time derivative of the
expression on
the left hand side of (\ref{82}) and to use the Wigner equation.
After this one has
\begin{eqnarray}
\int W^2(q,p;t)dq\, dp &=&
\biggl[\frac{a(t)}{a(0)} \biggr]^3 \biggl\{ \int W_0^2 (q,p)dq\,dp
\nonumber\\
& &-\frac{2\hbar}V \int^t_0 \biggl[\frac{a(0)}{a(\tau )}\biggr]^3 \cdot D(\tau
)
\cdot \biggl[\int \biggl(\frac{\partial W(q,p;\tau )}{\partial p} \biggr)^2
dq\,dp\biggr] d\tau \biggr\},
\label{83}
\end{eqnarray}
and (\ref{82}) follows from (\ref{83}).

If there is no diffusion, $D(t)\equiv 0$, for such a system, starting at $t=0$
from a pure state
\[
\int [W_0(q,p)]^2 dq\,dp =\frac{V}{2\pi\hbar} a(0)^3\
(<\frac{V}{2\pi\hbar} a(0)^3\mbox{ for a mixed state}),
\]
it is possible to follow the pure state in the course of time due to equality
\begin{equation}
\int W^2(q,p;t)dq\, dp =
\frac{V}{2\pi\hbar} a(t)^3\
(<\frac{V}{2\pi\hbar} a(t)^3\mbox{ for a mixed state}).
\label{84}
\end{equation}
However, diffusion ($D(t)\not\equiv 0$) smears the picture and one cannot
disinguish
pure and mixed states by inequality (\ref{82}).

For the Wigner function $W(q,\Pi ;t)$ (see (\ref{34})) relations (\ref{83}),
(\ref{82}) become
\begin{eqnarray}
\int W^2(q,\Pi ;t) dq\,d\Pi
&=&\int W^2_0 (q,\Pi )dq\,d\Pi
\nonumber\\
& &-\frac{2\hbar}V \int^t_0 a^6(\tau )D(\tau )
\biggl[\int \biggl(\frac{\partial W(q,\Pi ;\tau )}{\partial\Pi}
\biggr)^2 dq\,d\Pi \biggr]d\tau ,
\label{85}
\\
\int W^2(q,\Pi ;t) dq\,d\Pi
&\leq& \frac{V}{2\pi\hbar} .
\label{86}
\end{eqnarray}

\section{LARGE TIME ASYMPTOTICS}
\label{asymptotics}

If $H(t)\approx{\it Const}$ the scale factor $a(t)$ increases exponentially
with time and for large time the Wigner equation becomes
\begin{equation}
\frac{\partial W(q,p;t)}{\partial t}
= -p\frac{\partial W}{\partial q} +3H(t)\frac\partial{\partial p} (pW)
+ \frac{\partial U(q)}{\partial q}\cdot \frac{\partial W}{\partial p}
+\frac\hbar{V} D(t)\frac{\partial^2W}{\partial p^2}
\label{87}
\end{equation}
to order $a(t)^{-3}$, what follows from (\ref{39}).

The expectation value at large time $T$ for an operator $\hat A$ with Weyl
symbol $h(q,p)$ is
\begin{eqnarray}
\langle {\hat A}\rangle _T&=&\int h(q,p)W(q,p;T)dq\, dp
\nonumber\\
&=& \int\!{\bf E}\, h(\Phi_T(q,p,0),\dot\Phi_T(q,p,0))\,W_0(q,p)dq\, dp,
\label{88}
\end{eqnarray}
where the stochastic process $(\Phi_t,\dot\Phi_t)$ is governed now by
\begin{eqnarray}
d\Phi_t &=& \dot\Phi_t dt,
\nonumber\\
d\dot\Phi_t &=&
-[3H(t)\dot\Phi_t +\dot U(\Phi_t)]dt
+\biggl[ \frac{2\hbar}V D(t)\biggr]^{1/2}dw_t
\label{89}
\end{eqnarray}
with the initial condition $(\Phi_0,\dot\Phi_0)=(q,p)$.

Note that the stochastic differentials (\ref{89}) are equivalent to the quantum
  Langevin equations  (\ref{61})-- (\ref{63}) for the scalar field $\Phi(t)$ in
the large time limit (not for the beginning of the inflation if the potential
$U(q)$ satisffies the unequality  (\ref{40}).

\section{STATIONARY STATES FOR LARGE-SCALE INFLATION}
\label{stationary}

Stationary or equilibrium states for large-scale inflation are described by the
stationary Wigner function $W(q,p)=\lim_{t\to\infty}W(q,p;t)$ governed by the
Wigner equation
\begin{equation}
0 = -p\frac{\partial W}{\partial q} +3H(t)\frac\partial{\partial p} (pW)
+ \frac{\partial U(q)}{\partial q}\cdot \frac{\partial W}{\partial p}
+\frac\hbar{V} D(t)\frac{\partial^2W}{\partial p^2}
\label{90}
\end{equation}
for an arbitrary potential $U(q)$. This equation has time dependent
coefficients coming from the time dependence of the
scale factor.

For the stationary Wigner function $W(q,p)$ one can easily find the following
expectation values for the operators:
\begin{eqnarray}
\langle{\hat{\dot \Phi}}\rangle &=& 0,
\label{91}
\\
\langle{\hat{\dot U}}(\hat\Phi)\rangle &=& 0,
\label{92}
\\
\langle{\hat{\dot \Phi}}^2\rangle &=& \langle\hat\Phi{\hat{\dot
U}}(\hat\Phi)\rangle=
{\hbar D(t)\over {V\cdot3H(t)}}.
\label{93}
\end{eqnarray}

{}From the last equality one can see that the diffusion coefficient $D(t)$
should
be
\begin{equation}
D(t)=3 H(t)\,\sigma
\label{94}
\end{equation}
(at least for large time), where the constant $\sigma$ is defined by the choice
of stationary state:
\begin{equation}
\langle{\hat{\dot \Phi}}^2\rangle =\frac\hbar V\,\sigma.
\label{95}
\end{equation}

The stationary Wigner function is found to be
\begin{equation}
W(q,p) = N \exp \biggl\{ -\biggl(\frac{p^2}2+ U(q)\biggr)
V/(\hbar\sigma)\biggr\},
\label{96}
\end{equation}
where N is the normalization constant.

For a particular case, with the scale factor given by (\ref{10}) and potential
$U(q)=\omega^2q^2/2$, the Bunch-Davies vacuum is given by
\begin{equation}
\langle{\hat \Phi}^2\rangle =\frac{3 H^4\hbar}{8\pi^2\omega^2}=
\frac{\hbar H}{V2\pi\omega^2}
\label{97}
\end{equation}
which yields $\sigma=H/2\pi$ and $D=3H^2/2\pi$.

For this case the stationary solution is
\begin{equation}
W(q,p) = \frac{V\omega}{\hbar H}\exp \biggl\{ -(p^2+ \omega^2q^2)\frac{V\pi}
{\hbar H}\biggr\},\quad V=\frac{4\pi}{3H^3},
\label{98}
\end{equation}
where one must assume that
\begin{equation}
H \geq \omega/\pi
\label{99}
\end{equation}
for the inequality (\ref{44}) to be satisfied by the Wigner function
(\ref{98}).

\section{CONCLUSIONS}
\label{conclusions}

1. The appearance of the dissipation term $3H(t)(\partial/ \partial
p)(pW(q,p;t))$
in the Wigner equation, after transition from phase space $(\Phi,\Pi)$ to
$(\Phi,{\dot \Phi})$, supports the earlier result of Graziani \cite{9}
and Nakao, Nambu and Sasaki \cite{13}, that the large-scale inflation scalar
field behaves as a quantum one-dimensional dissipative system. Nevertheless,
this analogy is not complete: it is destroyed by a new commutation relation
\begin{equation}
[{\hat \Phi},{\hat{\dot \Phi}}] = \frac{i\hbar}{V} a(t)^{-3},
\label{100}
\end{equation}
where $a(t) $ is the scale factor in de~Sitter metric and it reflects
the expansion of the Universe (see (\ref{4})).

Comparing the Wigner equation (\ref{48}) for the large-scale inflation
(or more general case (\ref{36})--(\ref{37}))  with the
Wigner equation for a quantum one-dimensional linearly
damped unharmonic oscillator (see \cite{14,15,16}), one can see
that  the expansion of the Universe amplifies the role of the potential term,
which is a deviation from the harmonic oscillator potential. Explicitly this
amplification appears in formula (\ref{68}) for the expectation value of a
quantum operator and in formula (\ref{80}) for the Wigner function, where
function $A(t)$ (\ref{52}) has the factor $\int^t_0a(\tau )^3d\tau$ (without
expasion this factor would be 1). At the same time , as $t\to\infty$, the
commutation relation  (\ref{101}) leads to degeneration of the jump process, in
the quantum Langevin equations, into a continuous process (see
Section~\ref{asymptotics}).

2. As a consequence of our investigation we have  the following statement: for
the large-scale inflation scalar field the
asymptotic $t\to\infty$ is equal to the classical limit.

In the limit $\hbar\to 0$ the Wigner equation (\ref{36})--(\ref{37})
(or (\ref{48})) turns out to be
\begin{equation}
\frac{\partial W(q,p;t)}{\partial t}
=-p\frac{\partial W}{\partial q} +3H(t)\frac\partial{\partial p} (pW)
+\frac{\partial U(q)}{\partial q}  \cdot\frac{\partial W}{\partial p},
\label{101}
\end{equation}
and the corresponding Langevin equations are
\begin{eqnarray}
d\Phi_t &=& \dot\Phi_t dt,
\nonumber\\
d\dot\Phi_t &=& -[3H(t)\dot\Phi_t + U^\prime(\Phi_t)]dt.
\label{102}
\end{eqnarray}
Equations (\ref{102}) are just the classical deterministic equations of motion,
equivalent to the field equation  (\ref{11}).

However, (\ref{101}) is not the `classical limit' for the Wigner equation
 (\ref{36})-- (\ref{37}) (only the truncated Wigner equation of order $\hbar
_0$. In the classical limit the diffusion term in the Wigner equation
(\ref{36})-- (\ref{37}) is not proportional to $\hbar/V$ because, instead of
(\ref{95}) for a stationary state, one has
\begin{equation}
\langle{\dot\Phi^2}\rangle\,=\sigma_{cl}.
\label{103}
\end{equation}

The `classical limit' of the Wigner equation for the large-scale inflation
scalar field is
\begin{equation}
\frac{\partial W(q,p;t)}{\partial t}
=-p\frac{\partial W}{\partial q} +3H(t)\frac\partial{\partial p} (pW)
+\frac{\partial U(q)}{\partial q}  \cdot\frac{\partial W}{\partial p} +
D_{cl}(t)\frac{\partial^2 W}{\partial p^2}
\label{104}
\end{equation}
with $D_{cl} (t)=3H(t)\sigma_{cl}$. The classical stochastic Langevin equations
are
\begin{eqnarray}
d\Phi_t &=& \dot\Phi_t dt,
\nonumber\\
d\dot\Phi_t &=&-[3H(t)\dot\Phi_t + U^\prime(\Phi_t)]dt
+[2 D_{cl}(t)]^{1/2}dw_t.
\label{105}
\end{eqnarray}

One can see that that in the `classical limit' the Wigner equation and Langevin
equations ((\ref{104}) and (\ref{105})) are the same as those of the large time
asymptotic ( equations (\ref{87}) and (\ref{89})). For complete coincidence the
coefficient of the second derivative in the Wigner equations (\ref{87}) and
(\ref{104}) should be presented in the form
$3H(t)\langle{\dot\Phi^2}\rangle_{st}$, where the expectation value
$\langle{\dot\Phi^2}\rangle_{st}$ is taken on the corresponding stationary
state.

3. In this paper each large-scale region (for the coarse-graining procedure) is
considered as an independent quantum mechanical system. If, nevertheless, it is
necessary to take into account interaction with the environment, then for
linear
interaction the master equation for the `reduced' density operator $\hat\rho$
is
\begin{eqnarray}
\frac{\partial\hat\rho}{\partial t}
&=& \frac{V}{i\hbar} [\hat{\cal H} ,\hat\rho \,]
+\frac{\lambda (t)V}{4i\hbar} [\{\hat\Pi, \hat\Phi \},\hat\rho \,]
\nonumber\\
& &+\frac{\lambda (t)V}{2i\hbar} ([\hat\Phi ,\hat\rho \,\hat\Pi]
-[\hat\Pi ,\hat\rho\, \hat\Phi]) -\frac{V}{\hbar}  a(t)^6 D(t)
[\hat\Phi,[\hat\Phi,\hat\rho\, ]],
\label{106}
\end{eqnarray}
where $\lambda(t)$ is the dissipation coefficient (originated by interaction
with the environment), $\hat{\cal H}$ is an operator form for the Hamiltonian
(\ref{15}) of the system without dissipation, and $\{\,,\,\}$ stands for an
anti-commutator.

In phase space $(\Phi ,\dot\Phi )$ the Wigner equation corresponding to
(\ref{106}) is
\begin{eqnarray}  
\frac{\partial W(q,p;t)}{\partial t} &=&
-p\frac{\partial W}{\partial q} +[3H(t) +\lambda (t)]
\frac\partial{\partial p} (pW)
+\frac\hbar{V} D(t)\times
\nonumber\\
& &\times  \frac{\partial^2 W}{\partial p^2}
+\frac{V}{i\hbar} a(t)^3 \int^\infty_{-\infty} du\,
W(q,p-u;t) {\cal I}(q,u;t),
\label{107}
\end{eqnarray}
where ${\cal I}(q,u;t)$ is defined by (\ref{37}).

Equation (\ref{107}) is an equation of the same type as the Wigner equation
(\ref{36}). Thus, (\ref{107}) can be treated as the Wigner equation considered
in this paper.

4. Since in our consideration on phase space $(\Phi ,\dot\Phi )$ the scaling
$\hbar a(t)^{-3}/V$ depends already on $t$, it is not a problem to take into
account a time dependence of the coarse-graining volume $V(t)$. The scaling
becomes
$\hbar a(t)^{-3}/V(t)$, where $V(0)=\mbox{\it Const}\not= 0$ is assumed.

Therefor the result can be extended on an expanding FRW space-time. The
principal point in this case is the new time dependence of the canonical
momentum conjugate to $\Phi_X(t)$:
\begin{equation}
\Pi_\Omega(t) =  V(t)a(t)^3 \dot\Phi_X(t)
\label{108}
\end{equation}
and of the Hamiltonian:
\begin{equation}
{\cal H}_\Omega = \case 1/2 V(t)^{-1}a(t)^{-3} \Pi_\Omega^2 +
V(t)a(t)^3 U(\Phi_X )
\label{109}
\end{equation}
(compare with (\ref{20}) and (\ref{23})). Also the condition (\ref{24}), taken
for `approximatly' de~Sitter space, should be replaced by a condition
corresponding to each concrete FRW space-time.

\section*{Acknowledgments}

The support of the Royal Society via a postdoctoral fellowship
and the partial support of the Russian Academy of Sciences under grant No
94-02-06688 are acknowledged.
The author is very grateful to Dr. Ian G. Moss for constructive advice
and discussions, and to the mathematical physics group of the University of
Newcastle upon Tyne for a very fruitful and supportive atmosphere.

\end{document}